\documentclass[aps,preprint]{revtex4}

\usepackage{color}
\usepackage{graphicx}
\usepackage{amsmath}
\usepackage{amssymb}
\usepackage[]{hyperref}
\usepackage{bm}
\usepackage{ulem}

\usepackage{color} 

\begin{document}

\title{Quantum frequency conversion based on resonant four-wave mixing}

\author{Chin-Yao Cheng$^{1}$, Jia-Juan Lee$^{1}$, Zi-Yu Liu$^{1}$, Jiun-Shiuan Shiu$^{1}$, and Yong-Fan Chen$^{1,2,3}$}

\email{yfchen@mail.ncku.edu.tw}

\affiliation{$^1$Department of Physics, National Cheng Kung University, Tainan 70101, Taiwan \\
$^2$Center for Quantum Frontiers of Research \& Technology, Tainan 70101, Taiwan\\
$^3$Center for Quantum Technology, Hsinchu 30013, Taiwan
}



\begin{abstract}
Quantum frequency conversion (QFC), a critical technology in photonic quantum information science, requires that the quantum characteristics of the frequency-converted photon must be the same as the input photon except for the color. In nonlinear optics, the wave mixing effect far away from the resonance condition is often used to realize QFC because it can prevent the vacuum field reservoir from destroying the quantum state of the converted photon effectively. Under conditions far away from resonance, experiments typically require strong pump light to generate large nonlinear interactions to achieve high-efficiency QFC. However, strong pump light often generates additional noise photons through spontaneous Raman or parametric conversion processes. Herein, we theoretically study another efficient QFC scheme based on a resonant four-wave mixing system. Due to the effect of electromagnetically induced transparency (EIT), this resonant QFC scheme can greatly suppress vacuum field noise at low light levels; consequently, the converted photon can inherit the quantum state of the input photon with high fidelity. Our research demonstrates that if the conversion efficiency of the EIT-based QFC is close to 100\%, the wave function and quadrature variance of the converted photon are almost the same as the input probe photon. 
\end{abstract}


\pacs{42.50.Gy, 42.65.Ky, 03.67.-a, 32.80.Qk }


\maketitle


\newcommand{\FigOne}{
    \begin{figure}[t]
    \centering
    \includegraphics[width=11.0cm]{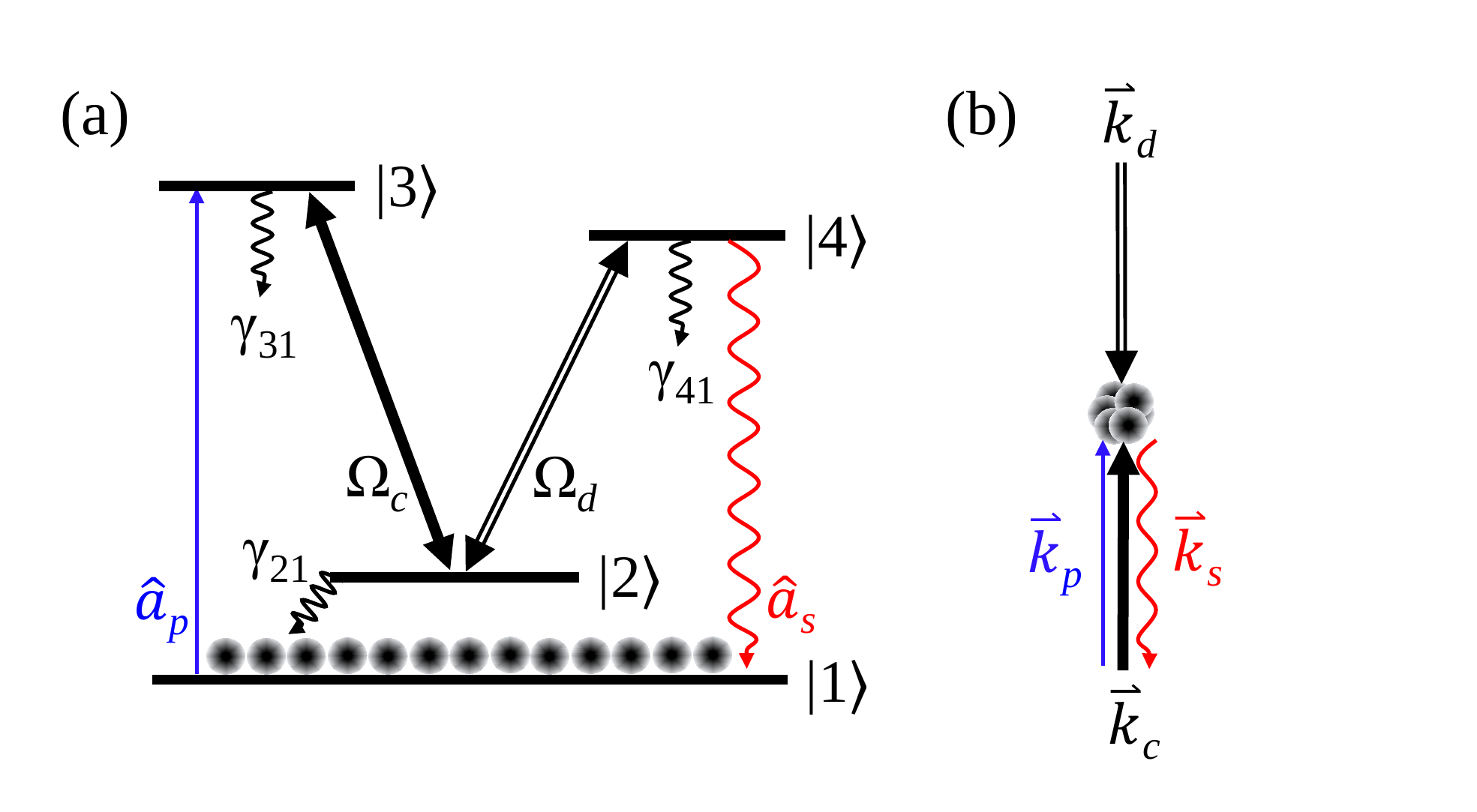}
    \caption{
(a) Energy level diagram of an EIT-based resonant FWM system. The probe and signal fields here are quantized. (b) Schematic of the propagation direction of each field vector in (a).}
    \end{figure}
}


\newcommand{\FigTwo}{
    \begin{figure}[t]
    \centering
    \includegraphics[width=10.6cm]{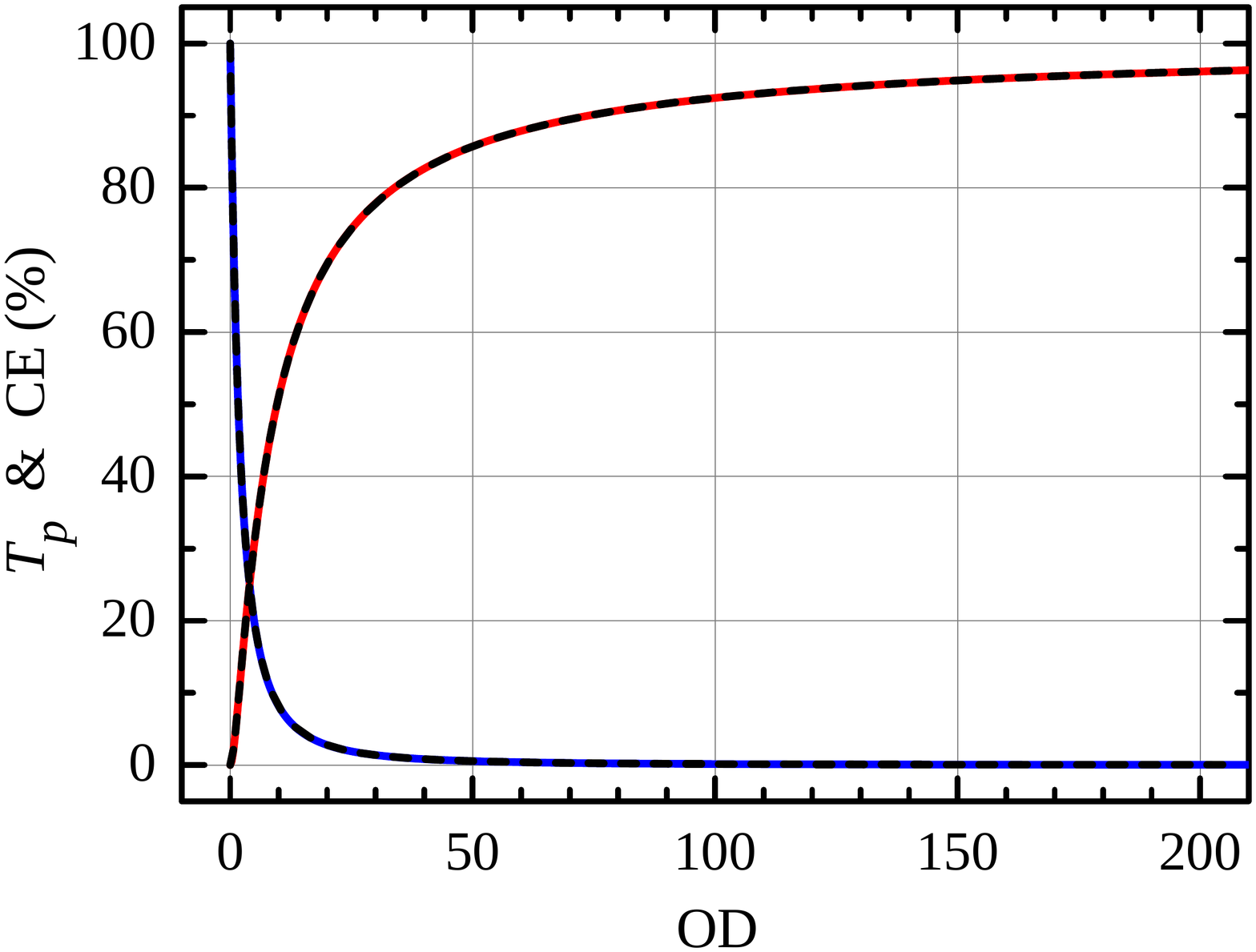}
    \caption{
Transmittance of the probe field (blue) and FWM efficiency of the converted signal field (red) versus the OD under the conditions of $\Omega_c=\Omega_d$, $\gamma_{31}=\gamma_{41}=\Gamma$, and $\gamma_{21}=0$. The solid and dashed lines indicate the theoretical curves calculated using the quantum and semiclassical models, respectively.}
    \end{figure}
}


\newcommand{\FigThree}{
    \begin{figure}[t]
    \centering
    \includegraphics[width=11.0cm]{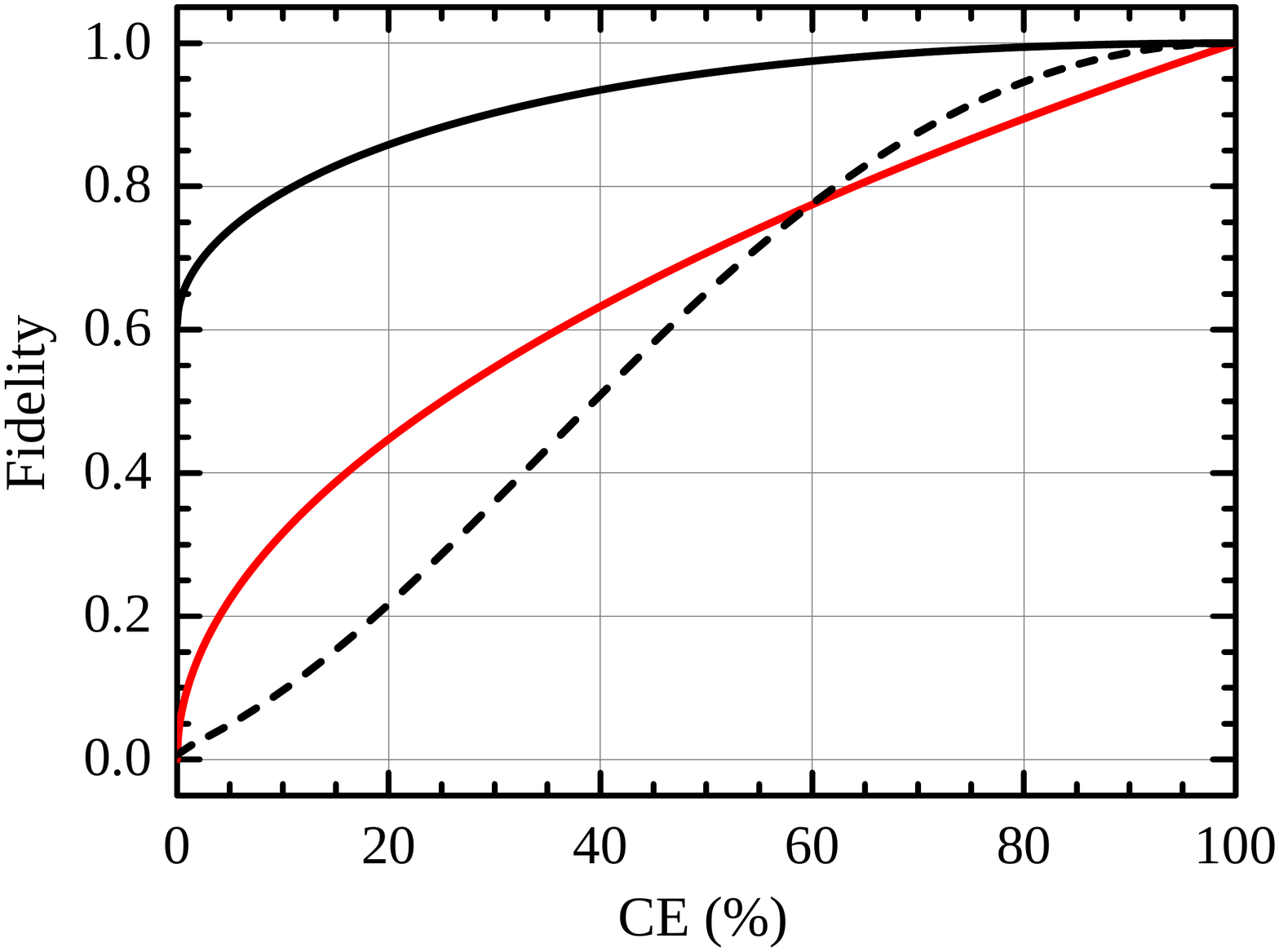}
    \caption{
The fidelity versus the CE of the EIT-based QFC. The black solid and dashed lines represent the theoretical predictions that the average photon number of the input probe field (coherent state) is 1 and 10, respectively. The red solid line is the theoretical curve when the input probe field is in the single-photon Fock state.}
    \end{figure}
}


\newcommand{\FigFour}{
    \begin{figure}[t]
    \centering
    \includegraphics[width=11.5cm]{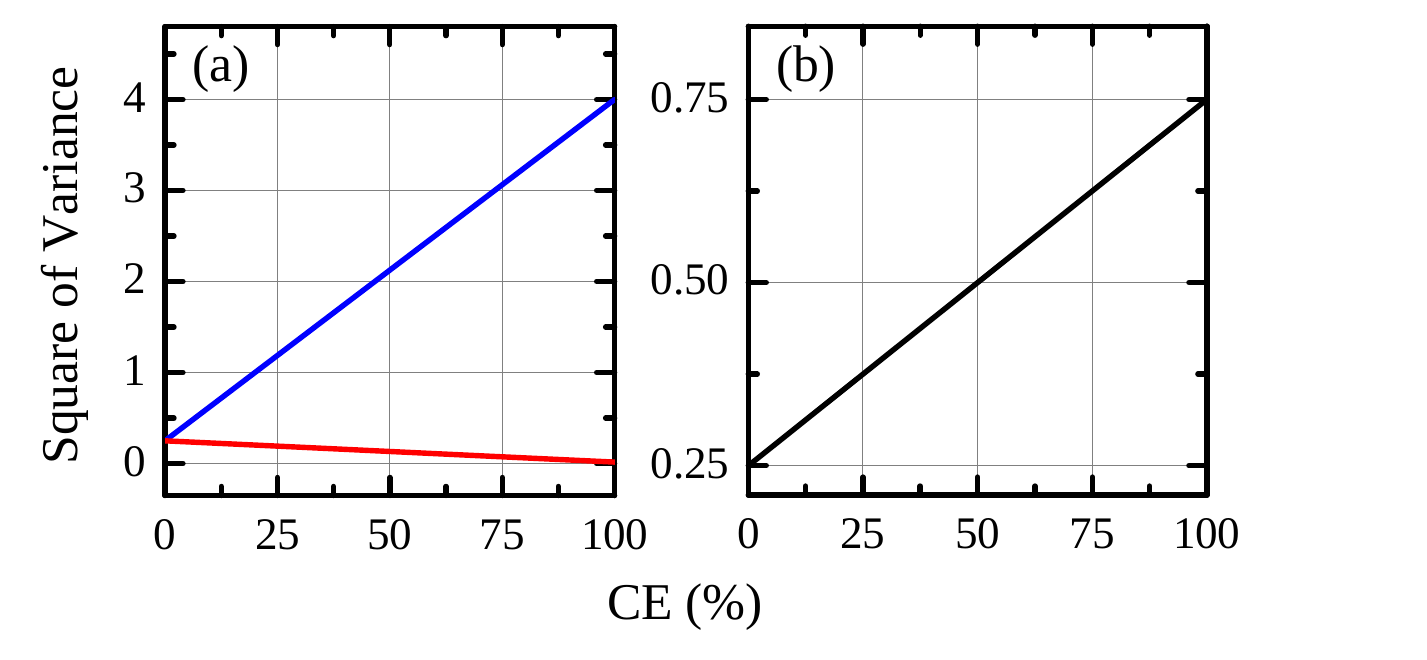}
    \caption{
Square quadrature variance of the converted signal field versus the CE of EIT-based QFC, where the input probe field is in (a) squeezed state (blue and red lines represent $\Delta X^{2}_s$ and $\Delta Y^{2}_s$, respectively) with $6$ dB squeezing, and (b) single-photon Fock state; $\Delta X^{2}_s$ and $\Delta Y^{2}_s$ are exactly the same regardless of CE in (b).}
    \end{figure}
}


\section{Introduction}

Quantum frequency conversion (QFC) can be used to not only connect photonic quantum devices with different frequency requirements but also generate multiple quantum states of photons; it thus plays a key role in long-distance quantum communication and effective optical quantum computing~\cite{QFC by Kumar, DLCZ, QC by Zeilinger, hadfield}. An ideal QFC only changes the color of the photonic qubit while leaving all other quantum properties unchanged. Frequency conversion experiments using a single-photon input have been effective when conducted in various nonlinear solid materials such as nonlinear crystals~\cite{SFG1, SFG4, Fisher, Allgaier, SFG5, Bock} and optical fibers~\cite{BS1, BS2, BS4}. Most of these QFC approaches based on solid materials have been realized under conditions far away from resonance because such conditions can effectively prevent the vacuum field reservoir from destroying the quantum state of the frequency-converted photons. 

When under conditions far away from  resonance, the strength of the interaction between light and matter is greatly reduced, and thus, experiments usually require a strong pump light to achieve high-efficiency QFC. However, under strong pump light conditions, additional noise photons are often generated due to spontaneous Raman or parametric conversion effects, which can cause difficulties in the practical application of QFC~\cite{Parametric Noise}. Although it is possible to reduce the pump power required for high-efficiency QFC by using waveguides or optical fibers, such a reduction causes the coupling loss of input photons, thus reducing  the overall efficiency of the QFC.

Another feasible mechanism through which to achieve high-efficiency QFC is to use a resonant four-wave mixing (FWM) system based on electromagnetically induced transparency (EIT)~\cite{Harris FWM, Fleischhauer EIT}. Many quantum applications based on EIT, including quantum memory~\cite{Hsiao QM, Laurat QM, Zhu QM}, photonic transistors~\cite{PTransister1, PTransister2, PTransister3}, optical phase gates~\cite{XPM1, XPM2, XPM3, XPM4}, and frequency beam splitters~\cite{Yu FBS}, have been proposed and demonstrated at the single-photon level because EIT can appreciably enhance the nonlinear interaction between photons and suppress vacuum field noise in free space. Some studies have confirmed that resonant FWM based on double-$\Lambda$ EIT can achieve extremely high conversion efficiency (CE)~\cite{Liu, Juo}. In this article, we theoretically study the quantum behavior of such an EIT-based FWM medium and discuss how the vacuum field reservoir affects the quantum properties of frequency-converted photons. Our research demonstrates that this resonant FWM system can be used for low-loss, high-fidelity QFC. If the CE of the resonant QFC is close to 100\%, the vacuum field noise is significantly suppressed and the quadrature variance of the converted photon is nearly identical to that of the input photon, regardless of the quantum state of the input photon.


\section{Theoretical Model}

We consider a four-level system with two ground states and two excited states, as shown in Fig. 1(a). The strong coupling field ($\Omega_c$ indicates its Rabi frequency) drives the transition between the ground state $|2\rangle$ and the excited state $|3\rangle$, thereby creating a transparent channel for the weak probe field driving the ground state $|1\rangle$ to the excited state $|3\rangle$ through the $\Lambda$-type EIT process. Under this EIT condition, the FWM process is induced by a strong driving field ($\Omega_d$), which drives the ground state $|2\rangle$ to the excited state $|4\rangle$, thereby converting the probe field into a signal field. In the FWM process, $\gamma_{31\left(41\right)}=\Gamma_{3\left(4\right)}+\gamma_{3\left(4\right)}$ represents the total coherence decay rate from the excited state $|3\rangle$ $(|4\rangle)$, where $\Gamma_{3(4)}$ and $\gamma_{3(4)}$ are the total spontaneous decay rate and the dephasing rate of the excited state $|3\rangle$ $(|4\rangle)$, respectively;  $\gamma_{21}$ is the dephasing rate between ground states $|1\rangle$ and $|2\rangle$. The probe and signal electromagnetic field operators are written in the quantized manner:

\begin{align}
 & \hat{E}^{(+)}_j=\sqrt{\frac{\hbar \omega_j}{2\epsilon_0 V}} \hat{a}_j(z,t)e^{-i\omega_jt+i\vec{k}_j\cdot\vec{z}},
\end{align}
where $V$ is the cross-sectional area of the medium multiplied by the length of the medium $L$. Permittivity in the vacuum is represented by $\epsilon_0$, $\hat{ a}_{j}$ is the slowly varying annihilation operator of the electrical fields, and the subscript $j$ can be applied to $p$ or $s$ to represent either the probe field or the signal field, respectively. Both the coupling field and the driving field maintain a semiclassical form because they are high light-level coherent states, which means that their quantum behavior is ignored here. It is worth noting that some theoretical studies have considered that both the weak probe field and the strong coupling field in the EIT medium are quantized. These studies indicate that as long as the dephasing rate between the ground states is zero, there is no quantum correlation between the probe field and the coupling field under steady-state conditions.~\cite{QEIT1, QEIT2}.

\FigOne 

Under the EIT condition of the weak probe field, the coupling field and the driving field propagating in the resonant FWM medium hardly lose energy, so their respective Rabi frequencies can be regarded as constants on the propagation $z$ axis. Therefore, the interaction Hamiltonian $\hat{H}$ for the four-level EIT-based FWM system is expressed as 
\begin{align}
\hat{H}=-\frac{\hbar N}{2 L}\int_{0}^L & \big[\Omega_{d}\tilde{\sigma}_{42}(z,t)+\Omega_{c}\tilde{\sigma}_{32}(z,t) \notag \\ & +2g_{p}\hat{a}_p(z,t)\tilde{\sigma}_{31}(z,t)\notag\\ & +2g_{s}\hat{a}_s(z,t)\tilde{\sigma}_{41}(z,t)+\text{H.c.}\big]dz
,\end{align}
where $N$ is the atomic number in the medium. $g_{p(s)}=\frac{d_{31(41)}\varepsilon_{p(s)}}{\hbar}$ denotes the coupling constant between the probe (signal) field and the medium. $d_{jk}$ is the dipole moment of the corresponding transition. $\varepsilon_{p(s)}=\sqrt{\frac{\hbar \omega_{p(s)}}{2\epsilon_0 V}}$ is  the electric filed of the single probe (signal) photon. The expression $\tilde{\sigma}_{jk}(z,t)$ represents a collective slowly varying atomic operator that obeys the Heisenberg--Langevin equation (HLE) between states $|j\rangle$ and $|k\rangle$, namely
\begin{align}
\frac{\partial}{\partial t}\tilde{\sigma}_{jk}=\frac{i}{\hbar}\big[\hat{H},\tilde{\sigma}_{jk}\big]-\frac{\gamma_{jk}}{2}\tilde{\sigma}_{jk}+\gamma^{sp}_{jk}+\tilde{F}_{jk},
\end{align}
where $\gamma^{sp}_{jk}$ and $\tilde{F}_{jk}$ represent the spontaneous decay rate and the Langevin noise operator, respectively.

Now, we consider a condition where all light fields are resonant in the FWM process. To obtain high FWM efficiency, the spontaneous emission loss in this resonant-type FWM scheme must be strongly suppressed. A simple solution is to arrange the applied laser fields to be configured backwards, as shown in Fig.~1(b)~\cite{Liu}. Because the driving field and the coupling field propagate in opposite directions, the direction of the generated signal field is also opposite to the input probe field in the backward FWM process.

When the probe field is weak, both the probe field and the signal field can be regarded as perturbation fields in the medium, and when considering the zero-order perturbation condition (i.e., the probe field is absent), all the population remains in the ground state $|1\rangle$, specifically $\langle\tilde{\sigma}^{(0)}_{11}\rangle =1$. To solve the first-order atomic operators, we substitute the zero-order results into the relevant first-order HLEs as follows:
\begin{align}
\frac{\partial}{\partial t}\tilde{\sigma}^{(1)}_{21} & =\tilde{F}_{21}-\frac{1}{2}\gamma_{21}\tilde{\sigma}^{(1)}_{21}-i\bigg[\frac{\Omega_c}{2}\tilde{\sigma}^{(1)}_{31}+\frac{\Omega_d}{2}\tilde{\sigma}^{(1)}_{41}\bigg], \\
\frac{\partial}{\partial t}\tilde{\sigma}^{(1)}_{31} & =\tilde{F}_{31}-\frac{1}{2}\gamma_{31}\tilde{\sigma}^{(1)}_{31}-i\bigg[g_p a^{\dagger}_p+\frac{\Omega^{\ast}_c}{2}\tilde{\sigma}^{(1)}_{21}\bigg], \\
\frac{\partial}{\partial t}\tilde{\sigma}^{(1)}_{41} & =\tilde{F}_{41}-\frac{1}{2}\gamma_{41}\tilde{\sigma}^{(1)}_{41}-i\bigg[g_s a^{\dagger}_s+\frac{\Omega^{\ast}_d}{2}\tilde{\sigma}^{(1)}_{21}\bigg].
\end{align}
To describe the probe and signal fields propagating in the FWM medium, we next use the Maxwell--Schr\"{o}dinger equations  as follows:
\begin{align}
\bigg(\frac{\partial}{\partial t}+c\frac{\partial}{\partial z}\bigg)\hat{a}^{\dagger}_p(z,t) & =-ig_p N \tilde{\sigma}^{(1)}_{31}(z,t), \\
\bigg(\frac{\partial}{\partial t}-c\frac{\partial}{\partial z}\bigg)\hat{a}^{\dagger}_s(z,t) & =-ig_s N \tilde{\sigma}^{(1)}_{41}(z,t).
\end{align}
By applying the Fourier transform $\hat{a}^{\dagger}(z,t)=\int \hat{a}^{\dagger}(z,\omega)e^{i\omega t}d\omega$ in Eqs. (4)--(8) and ignoring the relatively small term $\frac{i\omega}{c}$, Eqs. (7) and (8) can be rewritten as the coupled equations of $a^{\dagger}_p(z,\omega)$ and $a^{\dagger}_s(z,\omega)$ as follows:
\begin{align}
\frac{\partial}{\partial z}\hat{a}^{\dagger}_p+\Lambda_p \hat{a}^{\dagger}_p+\kappa_p \hat{a}^{\dagger}_s=\sum_{jk}\zeta^p_{jk}\tilde{f}_{jk},\\
\frac{\partial}{\partial z}\hat{a}^{\dagger}_s+\Lambda_s \hat{a}^{\dagger}_s+\kappa_s \hat{a}^{\dagger}_p=\sum_{jk}\zeta^s_{jk}\tilde{f}_{jk},
\end{align}
where $\tilde{f}_{jk}=\sqrt{\frac{N}{c}}\tilde{F}_{jk}(z,\omega)$ is defined as the renormalized Langevin noise~\cite{Kolchin}, $\Lambda_{p(s)}$ is the EIT profile coefficient, $\kappa_{p(s)}$ is the coupling coefficient of the probe (signal) transition, $\zeta^p_{jk}$ and $\zeta^s_{jk}$ are the coefficients of $\tilde{F}_{jk}$, and $\tilde{F}_{jk}$ denotes the Langevin noise operators of interest, $jk\in\lbrace21,31,41\rbrace$. To simplify the backward FWM model, we consider the conditions of $g_p=g_s=g$, $|\Omega_c|=|\Omega_d|=|\Omega|$, $\gamma_{31}=\Gamma_{3}=\gamma_{41}=\Gamma_{4}=\Gamma$, and $\gamma_{21}=0$. Notably, $\Gamma$ represents the spontaneous decay rate contributed by the vacuum field reservoir. According to the preceding conditions, the relevant parameters in Eqs. (9) and (10) can be obtained as follows:
\begin{align}
 & \Lambda_p=-\Lambda_s=\frac{\alpha\Gamma}{4L}(2i\Gamma\omega-4\omega^2+|\Omega|^2)/G(\omega),\\
 & \kappa_p=-\kappa_s=\frac{\alpha\Gamma}{4L}(-|\Omega|^2)/G(\omega),\\
 & \zeta^p_{21}=\sqrt{\frac{\alpha\Gamma}{4L}}(-2i\omega\Omega^{\ast}-\Gamma\Omega^{\ast})/G(\omega),\\
 & \zeta^p_{31}=\sqrt{\frac{\alpha\Gamma}{4L}}(4i\omega^2+2\Gamma\omega-i|\Omega|^2)/G(\omega),\\
 & \zeta^p_{41}=\sqrt{\frac{\alpha\Gamma}{4L}}(i|\Omega|^2)/G(\omega),\\
 & \zeta^s_{21}=\sqrt{\frac{\alpha\Gamma}{4L}}(\Gamma\Omega^{\ast}+2i\omega\Omega^{\ast})/G(\omega),\\
 & \zeta^s_{31}=\sqrt{\frac{\alpha\Gamma}{4L}}(-i|\Omega|^2)/G(\omega),\\
 & \zeta^s_{41}=\sqrt{\frac{\alpha\Gamma}{4L}}(-2\Gamma\omega-4i\omega^2+i|\Omega|^2)/G(\omega), 
\end{align}
where $G(\omega)=(\frac{1}{2}\Gamma+i\omega)(2i\Gamma\omega-4\omega^2+2|\Omega|^2)$. Note that in Eqs. (11)--(18),  we use the replacement of $\frac{g^2N}{c}=\frac{\alpha\Gamma}{4L}$, where $\alpha$ denotes the optical depth (OD) of the FWM medium. The general solutions of Eqs. (9) and (10) are given by
\begin{align}
\begin{bmatrix}
\hat{a}^{\dagger}_p(L)\\
\hat{a}^{\dagger}_s(L)
\end{bmatrix}
=
\begin{bmatrix}
A^{'} & B^{'}\\
C^{'} & D^{'}
\end{bmatrix}
\begin{bmatrix}
\hat{a}^{\dagger}_p(0)\\
\hat{a}^{\dagger}_s(0)
\end{bmatrix}
+\sum_{jk}\int_{0}^{L}e^{M(z-L)}
\begin{bmatrix}
\zeta^p_{jk}\\
\zeta^s_{jk}
\end{bmatrix}
\tilde{f}_{jk}dz,
\end{align}
where $\begin{bmatrix} A^{'} & B^{'} \\ C^{'} & D^{'} \end{bmatrix}=e^{-ML}$ and $M=\begin{bmatrix} \Lambda_p & \kappa_p \\ \kappa_s & \Lambda_s \end{bmatrix}$. Consider the boundary conditions $\hat{a}^{\dagger}_p(0,\omega)$ and $\hat{a}^{\dagger}_s(L,\omega)$ in the backward FWM system, and the expressions of $\hat{a}^{\dagger}_p(L,\omega)$ and $\hat{a}^{\dagger}_s(0,\omega)$ can be rewritten as
\begin{eqnarray}
\begin{bmatrix}
\hat{a}^{\dagger}_p(L)\\
\hat{a}^{\dagger}_s(0)
\end{bmatrix}
=
\begin{bmatrix}
A & B\\
C & D
\end{bmatrix}
\begin{bmatrix}
\hat{a}^{\dagger}_p(0)\\
\hat{a}^{\dagger}_s(L)
\end{bmatrix}
+\sum_{jk}\int_{0}^{L}
\begin{bmatrix}
P_{jk}\\
Q_{jk}
\end{bmatrix}
\tilde{f}_{jk}dz.
\end{eqnarray}
The matrix elements $A$, $B$, $C$, $D$, $P_{jk}$, and $Q_{jk}$ can be obtained by comparing the initial boundary value of field operators and noise terms between Eqs. (19) and (20) as follows:
\begin{gather}
\begin{bmatrix}
A & B\\
C & D
\end{bmatrix}
=
\begin{bmatrix}
A^{'}-\frac{B^{'}C^{'}}{D^{'}} & \frac{B^{'}}{D^{'}}\\
-\frac{C^{'}}{D^{'}} & \frac{1}{D^{'}} 
\end{bmatrix}
,\\
\begin{bmatrix}
P_{jk}\\
Q_{jk}
\end{bmatrix}
=
\begin{bmatrix}
1 & -\frac{B^{'}}{D^{'}}\\
0 & -\frac{1}{D^{'}} 
\end{bmatrix}
e^{M(z-L)}
\begin{bmatrix}
\zeta^p_{jk}\\
\zeta^s_{jk}
\end{bmatrix}
.
\end{gather}
Therefore, according to Eqs. (20)--(22), the creation operator of the probe field and the signal field in the frequency domain are respectively obtained as follows:
\begin{align}
\hat{a}^{\dagger}_p(L,\omega)= & A(\omega)\hat{a}^{\dagger}_p(0,\omega)+B(\omega)\hat{a}^{\dagger}_s(L,\omega) \notag\\
 & +\sum_{jk}\int_{0}^{L} P_{jk}(z,\omega)\tilde{f}_{jk}(z,\omega) dz, \\
\hat{a}^{\dagger}_s(0,\omega)= & C(\omega)\hat{a}^{\dagger}_p(0,\omega)+D(\omega)\hat{a}^{\dagger}_s(L,\omega) \notag\\
 & +\sum_{jk}\int_{0}^{L} Q_{jk}(z,\omega)\tilde{f}_{jk}(z,\omega) dz. 
\end{align}
The probe transmittance ($T_{p}$) and conversion efficiency (CE) of the backward FWM system are defined as the ratio of the mean photon number of the input probe field $\langle \hat{a}^{\dagger}_p(0,t_{0})\hat{a}_p(0,t_{0})\rangle =\textbf{n}_{p0}(0,t_{0})$ to the output probe field $\textbf{n}_{p}(L,t)$ and the converted signal field $\textbf{n}_{s}(0,t) $, respectively. Here, the noise correlation of vacuum reservoir is given by
\begin{align}
\langle\tilde{f}_{jk}(z,\omega)\tilde{f}_{k^{'}j{'}}(z^{'},\omega^{'})\rangle=\frac{L}{2\pi c}D_{jk,k^{'}j{'}}\delta(\omega-\omega^{'})\delta(z-z^{'}),
\end{align}
where $jk\in\lbrace21,31,41\rbrace$ is the subscript of the atomic operator $\tilde{\sigma}_{jk}$, and $k^{'}j^{'}\in\lbrace12,13,14\rbrace$ is its adjoint pair $\tilde{\sigma}_{k^{'}j^{'}}$. The parameter $D_{jk,k^{'}j{'}}$ represents the diffusion coefficient of the system, which can be obtained from the fluctuation\textendash dissipation theorem~\cite{Scully,Garrison}. Combine Eqs. (23) and (25) and use the inverse Fourier transform; then, the mean photon number of the output probe field in the time domain can be obtained as 
\begin{align}
\textbf{n}_p= & \langle \hat{a}^{\dagger}_p(L,t)\hat{a}_p(L,t)\rangle \notag\\
= &\int \int A(\omega)A^{\ast}(\omega^{'})\langle \hat{a}^{\dagger}_p(0,\omega)\hat{a}_p(0,\omega^{'})\rangle e^{i(\omega -\omega^{'})t} d\omega d\omega^{'} \notag\\
 &+\sum_{jk}\sum_{j^{'}k{'}}\int \int \frac{L}{2\pi c} P_{jk} D_{jk,k^{'}j{'}}P^{\ast}_{j^{'}k^{'}} dz d\omega.
\end{align}
Because the input probe (signal) field and the vacuum reservoir are statistically independent of each other, $\langle\hat{a}_{p(s)}\tilde{F}_{jk}\rangle$ and $\langle \hat{a}_{p(s)}^{\dagger}\tilde{F}_{jk}\rangle$ in Eq. (26) are zero. In addition, because the input signal field is regarded as a vacuum field in the theoretical model, $\langle \hat{a}_s(L,\omega)\rangle=\langle \hat{a}_s^{\dagger}(L,\omega)\rangle=\langle \hat{a}_s^{\dagger}(L,\omega)\hat{a}_s(L,\omega)\rangle=0$. Now, for simplicity, we consider the input probe field as a single-mode field and therefore require
\begin{align}
\langle \hat{a}^{\dagger}_p(0,\omega)\hat{a}_{p}(0,\omega^{'})\rangle=\delta(\omega)\delta(\omega^{'})\langle \hat{a}^{\dagger}_{p}(0,t_0)\hat{a}_{p}(0,t_0)\rangle.
\end{align}
Considering this single-mode case in the frequency domain is equivalent to assuming that the input probe field reaches a steady state condition in the time domain. Therefore, Eq.~(26) becomes 
\begin{align}
\textbf{n}_p
 = & |A_0|^2\textbf{n}_{p0} 
 +\sum_{jk}\sum_{j^{'}k{'}}\int \int \frac{L}{2\pi c} P_{jk} D_{jk,k^{'}j{'}}P^{\ast}_{j^{'}k^{'}} dz d\omega. 
\end{align}
For the single-mode case, $A_0$, $B_0$, $C_0,$  and $D_0$ denote the coefficients of $\omega=0$. In addition, the diffusion coefficient of the current FWM system can be obtained according to the following Einstein relation~\cite{Scully,Garrison}:
\begin{align}
D_{jk,k^{'}j^{'}}= & \frac{d}{dt}\langle\tilde{\sigma}_{jk}\tilde{\sigma}_{k^{'}j^{'}}\rangle-\left\langle\left[\frac{d}{dt}\tilde{\sigma}_{jk}-\tilde{F}_{jk}\right]\tilde{\sigma}_{k^{'}j^{'}}\right\rangle\notag\\ & -\left\langle\tilde{\sigma}_{jk}\left[\frac{d}{dt}\tilde{\sigma}_{k^{'}j^{'}}-\tilde{F}_{k^{'}j^{'}}\right]\right\rangle.
\end{align}
The obtained $D_{jk,k^{'}j^{'}}$ can be expressed as a matrix, as shown below:
\begin{eqnarray}
D_{jk,k^{'}j^{'}}=
\begin{bmatrix}
\frac{\Gamma}{2}\bigg(\langle\tilde{\sigma}_{44}\rangle+\langle\tilde{\sigma}_{33}\rangle \bigg) & 0 & 0\\
0 & 0 & 0\\
0 & 0 & 0
\end{bmatrix}
.
\end{eqnarray}
Under weak field perturbation conditions, the expectation values of the first-order atomic operators $\langle\tilde{\sigma}^{(1)}_{44}\rangle$ and $\langle\tilde{\sigma}^{(1)}_{33}\rangle$ are both zero; in addition, since the values of higher-order terms are very small and can be ignored, all matrix elements in $D_{jk,k^{'}j^{'}}$ are approximately zero. Therefore, the contribution of Langevin noise in Eq. (28) is zero. The transmittance of the probe field is given by
\begin{align}
T_p =\frac{\textbf{n}_p}{\textbf{n}_{p0}} =|A_0|^2=(\frac{4}{4+\alpha})^2.
\end{align}
Similarly, we can obtain the CE of the converted signal field as
\begin{align}
\texttt{CE} =\frac{\textbf{n}_s}{\textbf{n}_{p0}} =|C_0|^2=(\frac{\alpha}{4+\alpha})^2.
\end{align}
Figure 2 presents the theoretical curves of the probe transmittance and the FWM efficiency of the converted signal field versus the OD. The blue (probe) and red (signal) solid lines are calculated using Eqs. (31) and (32), respectively. The black dashed lines are plotted using the semiclassical model~\cite{Liu}; it is evident that the theoretical predictions of the quantum and semiclassical models are exactly the same.

\FigTwo


\section{State Of The Converted Photon}

We further study the quantum state of the converted signal photon in the backward resonant FWM system to check whether the quantum state is affected by the vacuum reservoir during the frequency conversion process. First, we write the density matrix of the output state of the FWM system as follows:
\begin{align}
\rho_f=U\rho_i U^{\dagger},
\end{align} 
where $\rho_i=\rho^{S}(L,t_0) \otimes \rho^{P}(0,t_0) \otimes \rho^{R}$ is the initial density matrix of the system and reservoir. The unitary matrix $U$ is the evolution operator of the combined system. The element of the density matrix of the converted signal field in the basis of Fock states is given by 
\begin{align}
 & \left\langle m\right|\rho^{S}(0,t)\left|n\right\rangle\notag \\ & =\left\langle m\right|{\rm {Tr}}_{P,R}\left[ U\rho_i U^{\dagger}\right] \left|n\right\rangle\notag \\ & ={\rm {Tr}}_{S}\left\lbrace \left|n\right\rangle\left\langle m\right|{\rm {Tr}}_{P,R}\left[ U\rho_i U^{\dagger}\right]\right\rbrace\notag \\ & ={\rm {Tr}}\left\lbrace (\left|n\right\rangle\left\langle m\right|\otimes I_{P}\otimes I_{R}) U\rho_i U^{\dagger} \right\rbrace\notag \\ & ={\rm {Tr}}\left\lbrace U^{\dagger} (\left|n\right\rangle\left\langle m\right|\otimes I_{P}\otimes I_{R}) U \rho_i \right\rbrace,
\end{align}
where ${\rm{Tr}}_{S}$ represents the trace over the degrees of freedom of the signal field, and $\rm {Tr}$ denotes the total trace of the combined system. $I_{P}$ and $I_{R}$ are the identity operators on the Hilbert space of the probe field and reservoir, respectively. Thus, the operator of the materix element $\rho^{S}_{mn}(0,t)$ is given by
\begin{align}
\hat{\rho}^{S}_{mn}(0,t) = U^{\dagger} (\left| n\right\rangle \left\langle m\right|\otimes I_{P} \otimes I_{R}) U.
\end{align}
According to the following equations obeyed by the creation and annihilation operators, namely
\begin{align}
\hat{a}\left|n\right\rangle=\frac{1}{\sqrt{n}} \left|n-1\right\rangle, & \\
\hat{a}^{\dagger}\left|n\right\rangle=\frac{1}{\sqrt{n+1}}\left|n+1\right\rangle, & \\
\left|0\right\rangle\left\langle 0\right|=\sum_{l=0}^{\infty} \frac{\left(-1\right)^l}{l!} (\hat{a}^{\dagger})^{l} (\hat{a})^{l},& 
\end{align}
$\hat{\rho}^{S}_{mn}(0,t)$ can be reformed as 
\begin{align}
 & U^{\dagger} (\left| n\right\rangle \left\langle m\right|\otimes I_{P} \otimes I_{R}) U \notag \\  & = U^{\dagger}\sum_{l=0}^{\infty} \mathcal{X}_{mnl} \left[\hat{a}_{s}^{\dagger}(L,t_0)\right]^{l+n} \left[\hat{a}_{s}(L,t_0)\right]^{l+m} U \notag \\ & = \sum_{l=0}^{\infty} \mathcal{X}_{mnl} \left[\hat{a}_{s}^{\dagger}(0,t)\right]^{l+n} \left[\hat{a}_{s}(0,t)\right]^{l+m},
\end{align}
where $\mathcal{X}_{mnl}$ denotes the coefficient $\frac{1}{\sqrt{m!n!}}\frac{\left(-1\right)^l}{l!}$ for simplicity. Combine Eqs. (24) and (39) and assume that the input probe is a single-mode field; then, the operator of the matrix element of the converted signal field is given by
\begin{align}
 & \hat{\rho}^{S}_{mn}(0,t) \notag\\ & =\sum_{l=0}^{\infty} \mathcal{X}_{mnl} \left[\hat{a}_{s}^{\dagger}(0,t)\right]^{l+n} \left[\hat{a}_{s}(0,t)\right]^{l+m} \notag\\ & =\sum_{l=0}^{\infty} \mathcal{X}_{mnl} \left[C_0 \hat{a}_{p}^{\dagger}(0,t)\right]^{l+n} \left[C_0^{\ast} \hat{a}_{p}(0,t)\right]^{l+m}.
\end{align}
Where the input probe field is the single-photon Fock state, the density matrix of the input probe photon is expressed as $\rho^{P}(0,t_{0})=\left|1\right\rangle\left\langle 1\right|$. According to Eq. (40), the element of the density matrix of the converted signal photon is given by
\begin{align}
 &  \rho^{S}_{mn}(0,t)={\rm {Tr}}\left[\hat{\rho}^{S}_{mn}(0,t)\rho_i\right]\notag \\ & ={{\rm {Tr}}_{P}}\left\lbrace\sum_{l=0}^{\infty} \mathcal{X}_{mnl} \left[C_0 \hat{a}_{p}^{\dagger}(0,t)\right]^{l+n} \left[C_0^{\ast} \hat{a}_{p}(0,t)\right]^{l+m}\left|1\right\rangle\left\langle 1\right|\right\rbrace.
\end{align}
Only two terms in the preceding equation are not zero: $\rho^{S}_{00}=1-\left|C_0\right|^2$ and $\rho^{S}_{11}=\left|C_0\right|^2$. Thus, the density matrix of the converted signal photon in the EIT-based FWM system is $\rho^{S}(0,t)=(1-\left|C_0\right|^2)\left|0\right\rangle\left\langle 0\right|+\left|C_0\right|^2\left|1\right\rangle\left\langle 1\right|$. According to Eq. (32), where the OD is large, the CE of the FWM system can reach nearly 100\%; thus, the quantum state of the converted signal photon is almost the same as that of the input probe photon, which is the main characteristic of QFC.

\FigThree

Next, we consider the case where the input probe field is in a coherent state. The density matrix of the input probe photon is expressed as $\rho^{P}(0,t_{0})=\left|\beta \right\rangle\left\langle \beta \right|$. The element of the density matrix of the converted signal photon is given by
\begin{align}
 &  \rho^{S}_{mn}(0,t)={\rm {Tr}}\left[\hat{\rho}^{S}_{mn}(0,t)\rho_i\right]\notag \\ & ={{\rm {Tr}}_{P}}\left\lbrace\sum_{l=0}^{\infty} \mathcal{X}_{mnl} \left[C_0 \hat{a}_{p}^{\dagger}(0,t)\right]^{l+n} \left[C_0^{\ast} \hat{a}_{p}(0,t)\right]^{l+m}\left|\beta\right\rangle\left\langle \beta\right|\right\rbrace\notag \\ & =\sum_{l=0}^{\infty} \mathcal{X}_{mnl} \left\langle \beta\right|\left[C_0 \hat{a}_{p}^{\dagger}(0,t)\right]^{l+n} \left[C_0^{\ast} \hat{a}_{p}(0,t)\right]^{l+m}\left|\beta\right\rangle \notag \\ & = \sum_{l=0}^{\infty} \frac{\left(C_0^{\ast}\beta\right)^{m}\left(C_0\beta^{\ast}\right)^{n}}{\sqrt{m!n!}}\frac{\left(-\left|C_0^{\ast}\beta\right|^2\right)^l}{l!}\notag \\ & = e^{-\left|C_0^{\ast}\beta\right|^2}\frac{\left(C_0^{\ast}\beta\right)^{m}\left(C_0\beta^{\ast}\right)^{n}}{\sqrt{m!n!}}.
\end{align}
This indicates that the converted signal photon inherits the coherent state characteristics of the input probe photon and its wave function is $\left|C_0^{\ast}\beta\right\rangle$. We further calculate the fidelity between the input probe photon state and the converted signal photon state according to the definition in~\cite{Nielsen}, where the example of the coherent state is $\left|\langle\beta|C_0^{\ast}\beta\rangle\right|$. Figure 3 is a plot of the fidelity of EIT-based QFC as a function of CE, as determined according to Eqs. (41) and (42). Where the input probe photon is the single-photon Fock state, the fidelity of the converted signal photon is equal to the square root of the CE, as shown by the red solid line in Fig.~3. Take the single-photon Fock state of the input probe field as an example, when the OD is $200$, the EIT-based QFC can achieve $96\%$ CE and $0.98$ fidelity.


\section{Quadrature Variance}

We next study the quadrature variance of the quantum state of the converted signal photon. The two Hermitian quadrature operators $X$ and $Y$ are respectively defined as
\begin{eqnarray}
X_s=\frac{1}{2}\left[\hat{a}_s(0,t)+\hat{a}_s^{\dagger}(0,t)\right],   \\ Y_s=\frac{1}{2i}\left[\hat{a}_s(0,t)-\hat{a}_s^{\dagger}(0,t)\right].
\end{eqnarray}
As in the previous section, the input probe photon is assumed to be a single-mode field in the frequency domain: $\hat{a}^{\dagger}_p(0,\omega)=\delta(\omega)\hat{a}^{\dagger}_{p}(0,t_0)$. Combine Eqs. (24) and (43) and assume that the input signal field is a vacuum state; then, the quadrature variance of the converted signal photon can be obtained as follows:
\begin{align}
 & \Delta X^2_s(0,t) = \langle X^2_s(0,t)\rangle -\langle X_s(0,t)\rangle ^{2} \notag\\
= & \frac{1}{4}\bigg\lbrace |C_0|^2\langle(\hat{a}_{p0}+\hat{a}^{\dagger}_{p0})^2\rangle + |D_0|^2\langle(\hat{a}_{s0}+\hat{a}^{\dagger}_{s0})^2\rangle \notag\\
 & - |C_0|^2\langle \hat{a}_{p0}+\hat{a}^{\dagger}_{p0}\rangle^2-|D_0|^2\langle \hat{a}_{s0}+\hat{a}^{\dagger}_{s0}\rangle^2 \notag\\
 & +\eta_{1}+\eta_{2}\bigg\rbrace.
\end{align}
Of these, $\hat{a}_{p0}=\hat{a}_p(0,t_0)$ and $\hat{a}_{s0}=\hat{a}_s(L,t_0)$ represent the input probe field and input signal field, respectively, and $\eta_{1}$ and $\eta_{2}$ represent the contribution of Langevin noise in the FWM system, as shown below:
\begin{align}
\eta_{1} = \sum_{jk}\sum_{j^{'}k{'}} \int \int \frac{L}{2\pi c} Q_{jk} D_{jk,k^{'}j^{'}}Q^{\ast}_{j^{'}k^{'}} dz d\omega,\\
\eta_{2} = \sum_{jk}\sum_{j^{'}k{'}} \int \int \frac{L}{2\pi c} Q^{\ast}_{j^{'}k^{'}} D_{k^{'}j^{'},jk}Q_{jk} dz d\omega.
\end{align}
Under the same conditions given in Eq.~(28), the noise term $\eta_{1}$ is zero. In addition, by using the commutation relation $ [\hat{a}_s(0,t),\hat{a}_s^{\dagger}(0,t)] = 1$, the noise term $\eta_{2}$ can be obtained as follows:
\begin{align}
\eta_{2}=1-|C_0|^2-|D_0|^2+\eta_{1}.
\end{align}
Therefore, Eq.~(45) is simplified as
\begin{align}
 & \Delta X^2_s(0,t) \notag\\
 = & \frac{1}{4}\bigg\lbrace |C_0|^2\big[4\Delta X^2_{p0}-1 \big]+|D_0|^2\big[4\Delta X^2_{s0}-1\big]+1 \bigg\rbrace \notag\\
= & |C_0|^2 \Delta X^2_{p0}+\frac{1}{4}(1-|C_0|^2),
\end{align}
where $\Delta X_{p0}$ and $\Delta X_{s0}$ respectively represent the quadrature variance of the input probe field and input signal field. Here, we use $\Delta X_{s0}=0.5$ because the input signal field is a vacuum state. Similarly,
\begin{align}
\Delta Y^2_s(0,t)=|C_0|^2 \Delta Y^2_{p0}+\frac{1}{4}(1-|C_0|^2).
\end{align}
In the same manner, the square quadrature variance of the output probe field can also be obtained:
\begin{align}
\Delta X^2_p(L,t) & =|A_0|^2 \Delta X^2_{p0}+\frac{1}{4}(1-|A_0|^2), \\
\Delta Y^2_p(L,t) & =|A_0|^2 \Delta Y^2_{p0}+\frac{1}{4}(1-|A_0|^2).
\end{align}
The square quadrature variance in Eqs. (49)--(52) can be divided into two parts. The first part contains $\Delta X^2_{p0}$ or $\Delta Y^2_{p0}$, indicating the quantum variance inherited from the input probe field. The second part is $\frac{1}{4}(1-|C_0|^2)$ or $\frac{1}{4}(1-|A_0|^2)$, which is only contributed from the vacuum reservoir. If $|C_0|^2 \rightarrow0$, the quantum characteristic inherited from the input probe field completely disappears. By contrast, if $|C_0|^2 \rightarrow 1$, meaning that CE approaches 100\%, the quadrature variance of the converted signal field is almost the same as the input probe field, thus exhibiting the characteristics of QFC.

\FigFour

We further use the input probe field in coherent, squeezed, and Fock states to calculate the quadrature variance of the converted signal field. If the input probe field is in a coherent state, regardless of CE, the quadrature variance of the converted signal field is 0.5. This result means that when the input probe field is in a coherent state, the quadrature variance of the converted signal field in the EIT-based FWM process remains unchanged. In this case, because the converted signal field undergoes a dissipation process caused by the vacuum reservoir, its wave function is $\left|C_0^{\ast}\beta\right\rangle$, as described in the previous section.

Figure 4(a) is a plot of the square quadrature variance of the converted signal field as a function of CE for the case where the input probe field is a squeezed state with 6 dB squeezing of $\Delta X_{p0}=2$ and $\Delta Y_{p0}=0.125$. When the CE is close to 100\%, the converted signal field completely inherits the quadrature variance of the input probe field. However, when CE approaches 0, the converted signal field returns to the coherent state contributed by the vacuum reservoir. Finally, where the input probe field is a single-photon Fock state, the change in the quadrature variance is similar to the case of the squeezed state. The difference is that $\Delta X^{2}_s$ and $\Delta Y^{2}_s$ are exactly the same in the case of the Fock state, as shown in Fig.~4(b)


\section{Conclusion}

In this article, we theoretically demonstrate that EIT-based resonant FWM can be used for high-fidelity QFC and its CE can nearly reach 100\% under ideal conditions. In addition, through theoretical analysis, we illustrate how the vacuum reservoir distorts the wave function and quadrature variance of the frequency-converted photon during the resonant FWM process. Our research shows that if the CE of EIT-based QFC is close to 100\%, then the wave function and quadrature variance of the converted photon are almost the same as those of the input probe photon. This high-fidelity QFC based on resonant FWM can be easily combined with EIT-related photon manipulation technology, meaning that it has the potential for application in optical quantum communication and optical quantum computing. 


\section*{ACKNOWLEDGMENTS}

We thank Po-Chen Kuo and Jhen-Dong Lin for their useful contributions to discussions. This work was supported by the Ministry of Science and Technology of Taiwan [grant number: 107-2112-M-006-008-MY3]. We also acknowledge the support of the Center for Quantum Technology in Taiwan.


\end{document}